\documentclass[%
 reprint,amsmath,amssymb,aps,
]{revtex4-1}

\usepackage{graphicx}% Include figure files
\usepackage{dcolumn}% Align table columns on decimal point
\usepackage{bm}% bold math
\usepackage{natbib}
\usepackage{hyperref}
\usepackage{phoenician}
\usepackage{hyperref}
\begin{document}

%\preprint{APS/123-QED}    

\title{On Axion's Effect on Propagation of Monochromatic Electromagnetic Wave Through Strong Magnetic Field}
\author{Mikhail Khankhasayev} 
\author{Carol Scarlett}%
\affiliation{Physics Department,Florida A{\&}M University}
\email{mikhail.khankhasayev@gmail.com}
\email{carol.scarlett@famu.edu}\date{\today}% It is always \today, today,

\begin{abstract}
A possibility of detecting the effect of photon-axion mixing in a cavity experiment is discussed. There are two photon-axion modes that acquire different indices of refraction and split in an inhomogeneous magnetic field. For a magnetic field inhomogeneous in the direction transverse to the light propagation an analytical solution is obtained both for the index of refraction and the beams' trajectories. In a cavity experiment, the beam splitting creates a bifurcation effect, which results in a decrease of the light intensity in the central region. Modulation of magnetic field can separate this effect from background by providing a narrow frequency range for any observed signal.  When one integrates this effect over time and accounts for bandwidth, the overall drop in FWHM intensity is of order $10^{-2}\%$. This is a very measurable effect.   
%\end{description}
\end{abstract}

\maketitle

%\tableofcontents
 
\section{\label{sec:level1}Introduction}

It is well known that magnetic fields change optical properties of the vacuum (permiability and pemittivity) due to photon polarization effects (see,e.g., \cite{Adler71}). The presence of axions creates additional optical effects via mixing of photons and axions in magnetic field. In addition to changes of the index of refraction the axion-photon interaction creates a Stern-Gerlach type effect of beam splitting in an inhomogeneous magnetic field (see,e.g. \cite{Raffelt88,Chelouche09}. The possible existence of axions have been the source of much theoretical and experimental interest. The axion was first introduced in Refs. \cite{Peccei77,Weiberg78} to solve the strong CP problem. In addition, as a light, weakly interacting and long-lived particle, the axion is considered as one the top candidates for explaining the dark matter (CDM) in the Universe (see, e.g., Refs. \cite{Wilcheck78,Sikivie85, Kim10}.  

In Ref.\cite{Chelouche09}, the possibility of detecting the mixing of axions and photons by studying the arrival times of radio pulses from highly magnetized pulsars was discussed. In the present letter we discuss the possibility of detecting the photon-axion mixing effects in a cavity experiment. The paper is organized as follows. In Section 2 we present a theoretical framework describing propagation of monochromatic light through a magnetized region. In Section 3 we present a discussion and quantitative estimates of measurable effects induced by axions. The final section summarizes our results and conclusions. 

\section{\label{sec:level1}Theory}
  
The description of the photon-axion interactions in an external magnetic field $\vec{B_e}$ is described by the Lagrangian density 
\begin{eqnarray}
L&=&-\frac{1}{4}F_{\mu \nu}F^{\mu \nu} + \frac{1}{2}({\partial_{\mu}a{\partial}^{\mu}a -{m_a}^2a^2)+\frac{1}{4}g_{\it a}{\it a}F_{\mu \nu}
{\tilde F}^{\mu \nu}} \nonumber \\ 
&+&\frac{\alpha ^2}{90{m_e}^4}\biggl[{\bigl(F_{\mu \nu}F^{\mu \nu}\bigr)}^2+\frac{7}{4}{\bigl(F_{\mu \nu}{\tilde F}^{\mu \nu}\bigr)}^2\biggr],  
\label{eq:L1}
\end{eqnarray}  
where $F_{\mu \nu}$ is the electromagnetic field tensor, ${\tilde F}^{\mu \nu}=\frac{1}{2}\epsilon _{\mu\nu \delta \gamma}F^{\mu \nu}$ is its dual tensor, $a$ and $m_a$ are the axion pseudoscalar field and its mass, respectively. The third term describes the photon-axion interaction, where $g_a$ is the axion-photon coupling constant. The fourth term describes the photon's dispersive effects induced by the external magnetic field \cite{Adler71}. 

In the presence of an external magnetic field it is convenient to present, following \cite{Adler71},the magnetic field as the sum, $\vec{B}\rightarrow \vec{B}+\vec{B}^e$.The photon-axion interaction term due to the external magnetic field is given by  
\begin{eqnarray} 
L_{int}=\frac{1}{4}g_{\it a}{\it a}F_{\mu \nu}{\tilde F}^{\mu \nu} = g_{\it a}{\it a}\bigl(\vec{B}^{e}\vec{E}\bigr).
\label{eq:interct1}
\end{eqnarray}

In the present paper we limit our considerations with a static (or slowly varying in time) magnetic field directed along the $x$-axis, i.e., $\vec{B}^e=\vec{e}_1B^e=(B^e,0,0)$, and  uniform along this direction. However, it is inhomogeneous in transverse direction (chosen as the y-axis) to the photon beam propagation. This conditions correspond to a typical setup in a cavity experiment when a laser beam propagates perpendicular to the magnetic field.    

According to Eq.(\ref{eq:interct1}), only the $x$-component of the photon's electric field contributes to the photon-axion interaction, and, for simplicity, we will consider linear, polarized light along the magnetic field, i.e., $\vec{E}=\vec{e}_1E_1=(E_1,0,0)=-\partial A_1/\partial t$, where $A_1$ is the corresponding component of the vector potential. In addition, we assume that the mirrors' surfaces are parallel to the external magnetic field. Therefore, during the reflections from mirror-to-mirror there will no change of the initial polarization of the light. Finally, the initial direction of the light beam is chosen to be along the $z$-axis. Since the photon-axion interaction is very weak, this will be the dominant direction for the propagation of light throughout the experiment.

To derive the system of equation describing axion-photon interaction in an external magnetic field let's neglect for a moment the photon-photon interaction term in Eq.(\ref{eq:L1}). The Lagrange equations in this case are the following: 
\begin{eqnarray}
&&\square A_1 +{g_a}B_e\frac{\partial a}{\partial t}=0 \nonumber \\
&&(\square-{m_a}^2)a -{g_a}B_e\frac{\partial A_1}{\partial t}=0,
\label{eq:lagrange1} 
\end{eqnarray}
where $\square ={\vec{\nabla}}^2- \frac{\partial ^2}{{\partial t}^2}$. 

To study propagation of photons and axions through the magnetic field it is naturally to look for the solutions in an eikonal form,
\begin{eqnarray}
A_1=\tilde{A} e^{i(\vec{k}\cdot\vec{r} -\omega t)}, \nonumber \\
a=\tilde{a}e^{i(\vec{k}\cdot\vec{r} -\omega t)},
\label{eq:ansatz}
\end{eqnarray}
where $\tilde{A}$ and $\tilde{a}$ denote the amplitudes of the photon and the axion correspondingly. By using this Anzatz we assume that both the photon and the axion beams propagate in space and pass through the external magnetic filed region together.        

Substituting these expressions in(\ref{eq:lagrange1}) we obtain,
\begin{eqnarray}
&& ({\omega}^2-{\vec{k}}^2)\tilde{A} -i{\omega}g_aB_e\tilde{a}=0, \nonumber \\
&& ({\omega}^2-\vec{k}^2-{m_a}^2)\tilde{a} +i{\omega}g_aB_e\tilde{A}=0.
\label{eq:lagrange2}
\end{eqnarray}
Here, we would like to note that in Refs.(\cite{Raffelt88,Chelouche09} the imaginary unit in the interaction term is missing. 

As it was shown in \cite{Adler71}, the photon-photon interaction in the presence of an external magnetic field creates dispersion effects which can be taken into account by introducing an effective photon's mass $Q_{\gamma}$ in Eq.(\ref{eq:lagrange2}, 
\begin{eqnarray}
&& ({\omega}^2-{\vec{k}}^2+Q_{\gamma})\tilde{A} -i{\omega}g_aB_e\tilde{a}=0, \nonumber \\
&& ({\omega}^2-\vec{k}^2 + Q_a)\tilde{a} +i{\omega}g_aB_e\tilde{A}=0.
\label{eq:lagrange3}
\end{eqnarray}
Here, $Q_a\equiv -{m_a}^2$ and 
\begin{eqnarray}
Q_{\gamma}={\omega}^2\frac{7\alpha}{45\pi}(\frac{B_e}{B_{crit}})^2,
\label{eq:photonmass}
\end{eqnarray}
where $B_{crit}=m_e^2\slash e\approx 4.4\times 10^{13}G$ is the critical magnetic field strength \cite{Raffelt88}. It is important to note that $Q_{\gamma}$ is a positive quantity. 
             
This system of equations can be written in a matrix form as 
\begin{eqnarray}
\left[({\omega}^2-{\vec{k}}^2)\hat{I}+ \left[ \begin{array}{cc}
Q_{\gamma}& -iQ_M\\
iQ_M & -m_a^2 \end{array}\right] \right]\left[ \begin{array}{c}
A\\
a \end{array}\right]=0, 
\label{eq:Matrix1}
\end{eqnarray} 
where $Q_M\equiv \omega g_aB^e$, and $\hat{I}$ is the unit $2\times 2$-matrix.   

Introducing $\lambda \equiv {\vec{k}}^2-{\omega}^2$, we can present this system as
\begin{eqnarray}
\left[ \begin{array}{cc}
Q_{\gamma}-\lambda & -iQ_M\\
iQ_M & Q_a -\lambda \end{array}\right]\left[ \begin{array}{c}
A\\
a \end{array}\right]=0.  
\label{eq:Matrix2}
\end{eqnarray}     

The eigenstates and eigenfuctions of the photon-axion system in the external magnetic field can be found by diagonalizing a Hermitian matrix 
\begin{eqnarray*}
\left[ \begin{array}{cc}
Q_{\gamma}-\lambda& -iQ_M\\
iQ_M  &Q_a -\lambda \end{array}\right].  
\label{eq:Matrix3}
\end{eqnarray*}      
The eigenvalues are give by 
\begin{eqnarray}
\lambda_{\pm}=\frac{1}{2}(Q_{\gamma}+Q_a)\pm \sqrt{\frac{1}{4}(Q_{\gamma}-Q_a)^2+Q_M^2},
\label{eq:disp2}
\end{eqnarray} 
or   
\begin{eqnarray}
k^2_{\pm}= {\omega}^2 +\frac{1}{2}(Q_{\gamma}+Q_a)\pm \sqrt{\frac{1}{4}(Q_{\gamma}-Q_a)^2+Q_M^2}.
\label{eq:disp3}
\end{eqnarray} 
The matrix $\hat{R}$ that mixes photons and axions in the magnetized area is given by   
\begin{eqnarray}
\hat{R}=\left[ \begin{array}{cc}
{\cos}{\phi}& i{\sin}{\phi}\\
i{\sin}{\phi}& {\cos}{\phi}\end{array}\right],  
\label{eq:Matrix4}
\end{eqnarray} 
where angle $\phi$ is determined by
\begin{eqnarray}
\frac{1}{2}{\tan}{2\phi}=\frac{Q_M}{Q_{\gamma}-Q_a}. 
\label{eq:mixangle}
\end{eqnarray} 
The photon-axion eigenstates $\tilde{A}^{\prime}$ and $\tilde{a}^{\prime}$ are given by 
\begin{eqnarray}
\tilde{A}^{\prime}={\cos}{\phi}\tilde{A} +i{\sin}{\phi}\tilde{a} \nonumber \\
\tilde{a}^{\prime}={\cos}{\phi}\tilde{a}+i{\sin}{\phi}\tilde{A}  
\label{eq:mixstates}
\end{eqnarray} 
in terms of pure photon $\tilde{A}$ and axion $\tilde{a}$ states. In the vacuum without magnetic field (let's label it below as Region 1)these amplitudes satisfy to the free Klein-Gordon equation.
   
\subsection{Birefringence}
A physical sense of the obtained eigenfunctions can be gained by considering the limit of $Q_M \rightarrow 0$ when the magnetic field is zero. In this limit, $\tilde{A}^{\prime}\rightarrow \\tilde{A}$ and $\tilde{a}^{\prime} \rightarrow \tilde{a}$. It means that the photon and axion entering a magnetized region (Region 2) will form coherent mixed photon-axion states propagating with the indices of refraction (see, Eq.(\ref{eq:disp3}), 
\begin{eqnarray}
n^2_{\pm}= 1 &+&\frac{1}{2\omega^2}(Q_{\gamma}+Q_a) \nonumber \\
&\pm& \frac{1}{\omega^2}\sqrt{\frac{1}{4}(Q_{\gamma}-Q_a)^2+Q_M^2},
\label{eq:refrin1}
\end{eqnarray} 
where $"+"$ stands for the $\gamma^\prime$ mode and $"-"$ for the $a^\prime$ mode.   

There is a special case when the mixing of photons and axions reaches it's maximum. It happens if in Eq. (\ref{eq:mixangle}),
\begin{eqnarray}
\frac{Q_M}{Q_{\gamma}-Q_a} \rightarrow \infty,
\label{eq:simmetry}
\end{eqnarray}
which means that $Q_M>>{Q_{\gamma}-Q_a}$, or $Q_{\gamma}=Q_a$. However, taking into account that $Q_{\gamma} >0$ and $Q_a =-m_a^2 <0$, the condition of $Q_{\gamma}=Q_a$ can never be met in the vacuum (see, also the discussion in Ref.\cite{Raffelt88}.      

The mixing angle in the maximum mixing case is $\phi=45\deg$, and the eigenstates are given by 
\begin{eqnarray}
\tilde{A}^{\prime}=\frac{1}{\sqrt{2}}(\tilde{A} +i\tilde{a})\nonumber \\
\tilde{a}^{\prime}=\frac{1}{\sqrt{2}}(\tilde{a} +i\tilde{A}).
\label{eq:refrinS}
\end{eqnarray}
The indices of refraction for the symmetric case are given by    
\begin{eqnarray}
n^2_{\pm}= 1 \pm \frac{Q_M}{\omega^2},
\label{eq:refrin3}
\end{eqnarray} 
or,  
\begin{eqnarray}
n_{\pm}\approx  1 \pm \beta, \nonumber \\
\label{eq:refrin4} 
\end{eqnarray}           
where $ \beta \equiv g_aB_e/2\omega$.  In the symmetric case the indices of refraction are linear in the magnetic filed strength.

The symmetry between the photons and axions in the magnetic field was discussed in \cite{Guendelman08} in the context of the duality of photons and axions in the zero axion's mass limit. The symmetry is exact under the following conditions only: (a) the axion's mass is zero, and (b) the photon polarization effects in a magnetic field are ignored. Otherwise, the duality between photons and axions is approximate.

\subsection{Beam splitting effect}
Calculated above the indices of refraction, Eqs.(\ref{eq:refrin1}), give the light propagation trajectories through the magnetized region using well-known equation (see, e.g., \cite{landau10}),
\begin{eqnarray}
\frac{d\vec{l}}{dl}=\frac{1}{n}\Bigl[{\vec{\nabla}n}-\vec{l}(\vec{l}\vec{\nabla}n)\Bigr].
\label{eq:optic2}
\end{eqnarray}   
where $\vec{l}$ is the unit tangent vector along the light propagation path, and $dl$ is the infinitesimal element of length along the light line. From this equation it follows that in the area described by a constant index of refraction corresponding to a uniform magnetic field the light will propagates along a straight line. Since the indices of refraction are different for the photon induced mode and the axion induced mode in accordance with Eq.(\ref{eq:refrin1})the light beam entering from Region 1 (free space) into Region 2 (magnetized area) will be separated into two beams. This effect disappears when the incident ray is normal to the surface separating Regions 1 and 2.  
 
In an inhomogeneous magnetic field the indices of refraction (\ref{eq:refrin1}) are not constant as well. In this case the light ray splits into two beams (even if the incident ray is normal to the surface). Let's assume, for simplicity, that the external magnetic field $\vec{B}^e$ is inhomogeneous in the transverse direction which is chosen as the $y$-axis. In this case the index of refraction in Eq.(\ref{eq:optic2}) will depend on $y$ variable only, and the system of equations is reduced to the equation  
\begin{eqnarray}
\frac{dl_y}{dl}=\frac{1}{n}\frac{\partial n}{\partial y}\left(1-l_y^2\right), 
\label{eq:optic3}
\end{eqnarray}   
since the $z$-component of the unit tangent vector $\vec{l}$ can be found from the condition, $l_y^2+ l_z^2=1$.

Represent the beam's trajectory as $z=z(y)$ it is easy to find the following exact solution: 
\begin{eqnarray}
l_y=\Biggl\{\begin{array}{cc}
\sqrt{1-C\slash n^2},& n>1,\\
\\
-\sqrt{1-C{n^2}},& n<1.\end{array}
\label{eq:optic4}
\end{eqnarray}
Here, $C$ is an arbitrary constant to be determined by the initial condition $l_y=l_y(y_o,z_0)$. For example, if the light ray enters Region 2 at point $P=(y_0,z_0)$ normally, i.e., $l_y=0$, then   
\begin{eqnarray}
l_y=\Biggl\{\begin{array}{cc}
\sqrt{1-\frac{n_0^2}{n^2}}& n>n_0,\\
\\
-\sqrt{1-\frac{n^2}{n_0^2}}& n<n_0,\end{array}
\label{eq:optic4}
\end{eqnarray}
where $n_0 = n(y_0,z_0)$.      
 
The trajectories of photon-axion modes  in the magnetized region are determined by the following integrals: 
\begin{eqnarray}
z(y)=\int_{y_0}^y{\frac{n_0 dy}{\sqrt{n^2(y)-n^2_0}}+z_0}, \, n>n_0;\nonumber \\
z(y)=-\int_{y_0}^y{\frac{n(y)dy}{\sqrt{n^2_0-n^2(y)}}+z_0}, \, n<n_0. 
\label{eq:optic6}
\end{eqnarray}
Equations (\ref{eq:optic4}) and (\ref{eq:optic6}) present the complete solution to the problem of determining both the angle of refraction and the path of the light beam passing through the magnetized area.

If the index of refraction has a linear dependence on $y$, i.e.,  
\begin{equation}
n(y)=n_0+b(y-y_0),
\label{eq:linear1}
\end{equation}
where $b\equiv \partial n\slash \partial y$ and $n_0=n$ are taken at $y=y_0$, the integrals (\ref{eq:optic6})can be calculated analytically:
\begin{eqnarray}
z(y)&=&z_0 + \frac{n_0}{b}\ln\frac{2n^2}{n_0(n+n_0)},\, n>n_0,
\label{eq:path1}
\end{eqnarray}
and 
\begin{eqnarray}
z(y)=z_0-\frac{1}{b}\sqrt{n_0^2-n^2}, \, n<n_0.
\label{eq:path2}
\end{eqnarray}

\section{\label{sec:level1}Quantitative Estimates of Measurable Effects Induced by Axions} 

In this section we limit our analysis with the light beam entering Region 2 normal to the surface dividing Regions 1 and 2 at the point $y=0,z=0$. To separate the effect caused by the inhomogeneity of the magnetic field it is convenient to present the index of refraction in the following form   
\begin{eqnarray}
n= n_0 \pm \delta n, 
\label{eq:est1}
\end{eqnarray}
where $n_0=n(0,0)$ is the constant piece of the index of refraction; $\delta n$ is its variable part; $"+"$ corresponds to the $\gamma^{\prime}$ mode ($n>1$) ,and "-" to the $a^{\prime}$ mode ($n<1$), respectively. One can see that the constant part $n_0$ is different for $\gamma^{\prime}$ and $a^{\prime}$ modes.
  
Due to an extremely weak coupling ofthe photon to the axion, $g_a \sim 10^{-10} - 10^{-14} GeV^{-1}$, the axion induced effects are extremely small, i.e., $\delta n \ll n_0$.  Using Eq.(\ref{eq:optic4}) one can obtain the following lowest order approximation for $l_x$:  
\begin{eqnarray}
l_y\approx\pm \sqrt{2\delta n\slash n_0}.   
\label{eq:est2} 
\end{eqnarray}
For a linear inhomogeneity (\ref{eq:linear1}) this equation reads as 
\begin{eqnarray}
l_y\approx\pm \sqrt{2by\slash n_0}.   
\label{eq:est3}
\end{eqnarray}

For the same reason the shift of the light beam in the $y$-direction, $y=l_yL$ is much smaller the corresponding distance ($L$) traveled by light in the $z$-direction. Using Eqs.(\ref{eq:path1}, \ref{eq:path2}) one can find the following relationship between these two quantities     
\begin{eqnarray}
y\approx\frac{b}{2n_0}L^2.   
\label{eq:est4}
\end{eqnarray} 
Substituting this equation into (\ref{eq:est2}) we obtain
\begin{eqnarray}
l_y =\sin\theta\ \Rightarrow  \theta =\pm bL\slash n_0,   
\label{eq:est5}
\end{eqnarray}
in terms of the angle of refraction $\theta$. The splitting angle $\Delta \theta =\theta_{+}-\theta_{-}$ between the photon and axion induced modes is then given by  
\begin{eqnarray}
\Delta\theta = 2bL.
\label{eq:est7}
\end{eqnarray}
This formula is derived under the assumption that the magnetic field is inhomogeneous linearly in the transverse (y) direction,i.e., 
$$      
B^e(y)=B^e_0 + B^e_1 (y-y_0), 
$$
where $B^e_1 \equiv {\partial B^e \slash \partial y\vert} _{y=0}$. 

For the case of maximum mixing of photons and axions determined by Eq.(\ref{eq:refrin4}) we obtain:
\begin{eqnarray}
n_{\pm}\approx  1 \pm \beta_0,
\label{eq:refrin4}
\end{eqnarray}            
where $ \beta_0 \equiv g_aB^e_0/2\omega$, $B^e_0$ is the constant component of the external magnetic field, giving    
\begin{eqnarray}
{\theta}_{\pm} \approx \pm \frac{g_aB^e_1L}{2\omega n_0}.   
\label{eq:est6}
\end{eqnarray}
Therefore, the splitting angle $\Delta \theta =\theta_{+}-\theta_{-}$ between two modes is  
\begin{eqnarray}
\Delta\theta \approx \frac{g_aB^e_1L}{\omega}.
\label{eq:est7}
\end{eqnarray}

The obtained formula similar to that of Ref.\cite{Chelouche09}:$\Delta\theta \approx g_af_GB^e\slash\omega$, where $f_G$ is so-called geometric factor. Comparing these formulas, one obtains 
\begin{eqnarray}
f_G=(B^e_1\slash B^e)L=\frac{\partial\ln B^e}{\partial y}L,
\end{eqnarray}
where $L$ is the distance travel by the photon beam along the $z$-axis in the magnetic field with gradient $B^e_1$ along the $y$-axis.

\begin{figure}[htb]
\includegraphics[scale=0.25]{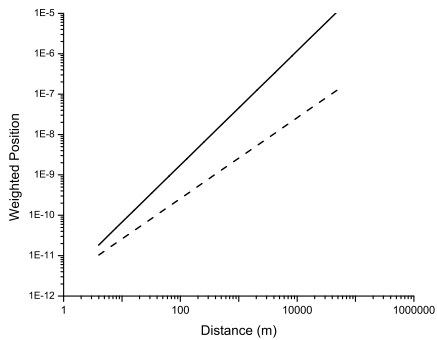}% Here is how to import EPS art 
\caption{\label{fig:bifurc_b} The spreading of rays for a bifurcating distribution (solid line) compared to that of a linear propagation of two beams (dashed line).} 
\end{figure}

In Ref.\cite{Chelouche09}, the authors discussed the possibility to detecting axions by studying radio pulses arriving from magnetars with magnetic fields of the order of $10^{16}G$ and the radii $\sim 10^5m$. To estimate the effect the authors used $f_G=0.1$, which translates to the magnetic field gradient of $B^e_1 \sim 10^{11}G\slash m$. This strong variation of the magnetic field over distances comparable to radio wave lengths can create an observable splitting angles $\sim 10^{-2}rad$.    
In laboratory experiments the typical strength of magnetic field is $\sim 10^4 - 10^5 G$, the size of the magnetic field region is $\sim 1 m$, and the magnetic field gradient is $\sim 10^6 G\slash m$. The geometric factor in this case is $f_G\approx 50$. Assuming a photon's wave length $\sim 10^{-6} m$ and using (\ref{eq:est7}) we can obtain the the following expression for the splitting angle $\Delta\theta \sim 10^{-5}g_a$ as a function of $g_a$ measured in $GeV^{-1}$. If we take the CAST limit \cite{Andriamonje07}  for the upper value of $g_a$, which is $10^{-10}$ then we predict a splitting of $\sim 10^{-15} rads$. 
One way to build up the effect, see Eq.(\ref{eq:est7}), is to increase the length L of the field region. In a cavity experiment, allowing the light to bounce back and forth between two mirrors will effectively increase the magnetic field length up to about $10^5 m$, where mirror reflectivity reaches its limit. Multiplying our effect by this total distance would give a splitting angle of $\sim 10^{-10} rad$.  HOWEVER, for a cavity experiment, reflection from each mirror destroys the axion-photon modes. After the first bounce, two beams reenter the magnetic field and divide giving four new beams, then eight, etc.  Importantly, the polarization of each new beam remains unchanged.  The effect is a bifurcation at the face of each mirror and in a distance of only $10^4 m$, less then mirror limitations, results in an average separation of the beam intensity of $\sim 10^{-9} rad$.  Assume for simplicity that two planar mirrors enclose a region of the magnetic field gradient that is $\sim 1m$ in length.
\begin{figure}[htb]
\includegraphics[scale=0.5]{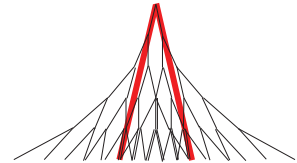}% Here is how to import EPS art 
\caption{\label{fig:bifurc_a}Bifurcation diagram: thick black lines represent the bifurcating rays, two bold lines represent a linear split of a beam}
\end{figure}
Figure \ref{fig:bifurc_a}shows a sketch of a bifurcating distribution for only a few bounces.  The two thick, straight lines show the linear solution for comparison.  
The splitting angle has been exaggerated to illustrate how much more rapidly a bifurcation diverges than a linear separation.  It can be shown that the overall density of rays for the bifurcated distribution expands at a rate of $Az^{\sqrt{2}}$. 
Figure \ref{fig:bifurc_b} uses this formula to show that after $10^4$ bounces assuming the above typical cavity experimental parameters, $( \theta \sim 10^-15 )$, the rays are shifted away from center by as much as $10^{-9}m$.  
In Figure \ref{fig:bifurc_b}, Weighted Position refers to the position of the new rays weighted by the fraction of initial light intensity each represents. Summing the weighted position gives the average separation of the light intensity from the central region.  For a Gaussian beam, energy spreads away from the center and is, for such a small shift, roughly linear giving rise to a drop in the FWHM of $\sim 10^{-9}E_0$, where $E_0$ is the initial energy.  Furthermore, we can suppress backgrounds through modulation of the external field.  When one integrates this effect over time and accounts for bandwidth, the overall drop in FWHM intensity is of order $10^{-4}$, which is a very measurable effect. 
A more detailed computer simulation results and computer code for the analyses of the bifurcation process that incorporates both the realistic magnetic field configuration (e.g., for a quadrupole magnet), shape and position of  mirrors, etc. will be presented in the subsequent paper.  

\section{\label{sec:level1}Conclusion}  
The present paper was focused on exploring a possibility of observing the mixing of axion and photons in a cavity experiment with an inhomogeneous magnetic field. The discussion was heavily based on the results of Refs. (\cite{Raffelt88,Chelouche09}). The presented formalism provides a consistent way of calculations the effects related to the mixing of photons and axions passing through the magnetized area. Eq.(\ref{eq:lagrange2}) and the obtained mixing matrix (\ref{eq:Matrix4}) correctly allow one to analyze the photon-axion mixing effects at arbitrary mixing angle providing a correct mixing phase and continuous transition to the symmetric mixing case corresponding $\phi = 45^o$. For an external magnetic field inhomogeneous in the direction transverse to the incident light we obtained an analytical solution, Eqs.(\ref{eq:optic4}) and (\ref{eq:optic6}), for the index of refraction and the beam's path. Assuming a linear inhomogeneity of the index of refraction we obtained expressions, Eq.(\ref{eq:path1}) and (\ref{eq:path2}), for the trajectories of the photon and axion induced modes in a magnetized area. The formulas provide an effective tool for analyzing both qualitatively and quantitatively propagation of light in a cavity experiment since they can be used to calculate both the direction and the point of entry into the magnetic field. Axions in the presence of external magnetic fields change the optical properties of the vacuum in several ways. First, the axion-photon mixing creates two indices of refraction, Eq.({\ref{eq:refrin1}) for the photon- and axion-induced modes which creates the effect of birefringence. In addition, in an inhomogeneous field these two mixed photon-axion modes are subjected splitting - a Stern-Gerlach type effect. The splitting effect creates a bifurcation effect for the light beam bouncing between two (or more) mirrors in a cavity type experiment. It is shown that bifurcation strongly increases the signature of the axion induced effects.   

\begin{acknowledgements}
This work was made possible in part by an NSF EArly concept Grant for Exploratory Research (EAGER) and with support from the Florida Scholars Boost Program.  These institutions are invaluable to scientific exploration and to future efforts to understand the nature of Cold Dark Matter in the universe.
\end{acknowledgements}

\nocite{*}

\bibliography{AxmF1}% Produces the bibliography via BibTeX.

\end{document}